\documentclass[showpacs,epsf,twocolumn]{revtex4-1}
\usepackage{float}
\usepackage{color,graphicx}
\usepackage{hyperref}
\begin{document}

\title{ Anisotropic Fermi surface probed by the de Haas-van Alphen oscillation in proposed Dirac Semimetal TaSb$_{2}$}

\author{Arnab Pariari, Ratnadwip Singha, Shubhankar Roy, Biswarup Satpati, Prabhat Mandal}

\affiliation{Saha Institute of Nuclear Physics, HBNI, 1/AF Bidhannagar, Kolkata 700 064, India}
\date{\today}

\begin{abstract}
TaSb$_{2}$ has been predicted theoretically and proposed through magnetotransport experiment to be a topological semimetal. In earlier reports, the Shubnikov-de Haas oscillation has been analyzed to probe the Fermi surface, with magnetic field along a particular crystallographic axis only. By employing a sample rotator, we  reveal highly anisotropic transverse magnetoresistance by rotating the magnetic field along different crystallographic directions. To probe the anisotropy in the Fermi surface, we have performed magnetization measurements and detected strong de Haas-van Alphen (dHvA) oscillations for the magnetic field applied along \textbf{b} and \textbf{c} axes as well as perpendicular to \textbf{bc} plane of the crystals. Three Fermi pockets have been identified by analyzing the dHvA oscillations. Hall measurement reveals electron as the only charge carrier, i.e., all the three Fermi pockets are electron type. With the application of magnetic field along different crystal directions, the cross sectional areas of the Fermi pockets have been found significantly different. Other physical parameters, such as the effective mass of the charge carrier and Fermi velocity have also been calculated using the Lifshitz-Kosevich formula.\\
\end{abstract}
\maketitle
\section{Introduction}
Inclusion of topology in electronic band structure has opened-up a new era in condensed matter research. Over the past one decade, the topological insulating phase of matter has emerged through continuous evolution from two-dimensional quantum spin Hall state to three-dimensional (3D) topological insulator \cite{hasan,bansil}. The bulk of this electronic phase of matter exhibits a band gap like an ordinary insulator whereas the edge or surface hosts symmetry protected highly conducting states. 3D Dirac and Weyl materials are the most recent discovery on quantum phase of matter, described as topological semimetal \cite{wang1,wang2,shin,liu,liu1,lv,bian}. Unlike topological insulators, these systems exhibit semimetallic bulk with linearly dispersing excitation and their surface state is topology protected Fermi arc. Due to the unique band topology, they show different exotic electronic properties of technological and fundamental interest. On the other hand, the nature and geometry of the Fermi surface can also modify the electronic transport significantly. Such as, the anisotropic magnetoresistance has been ascribed to the anisotropic nature of the Fermi surface \cite{kwang,zhao,ali1}. Thus, it is important to acquire the knowledge of Fermi surface to explain different electronic properties of a material.\\

Without taking into account the role of spin-orbit coupling, TaSb$_{2}$ has been proposed to be a topological semimetal. Upon inclusion of spin-orbit coupling, however, a topological gap opens-up at each band crossing point \cite{li,xu,zwang,yli}. This leads to the possibility of suppressed linear electronic dispersion in TaSb$_{2}$. Only the transport experiments have been performed so far to establish the three-dimensional Dirac fermionic excitation through the observation of negative longitudinal magnetoresistance (LMR) and detection of non-trivial $\pi$ Berry's phase in Landau level index plot \cite{li,yli}. Although the negative LMR has been established as a key signature of chiral anomaly, it can also appear due to several other reasons \cite{jing,kiku}. The existence of three-dimensional Dirac fermionic excitation in a material will be more convincing, if the observed LMR follows the typical magnetic field and temperature dependence, which have been predicted for chiral anomaly \cite{niel,quing,pariari}. The Landau level index plot for the detection of non-trivial Berry's phase is  reliable, only when one oscillation frequency is present instead of multiple frequencies in quantum oscillation. So, further investigations are needed to unambiguously establish the non-trivial band topology in TaSb$_{2}$. Apart from this unconventional nature of electronic band structure, large magnetoresistance and the presence of two or three Fermi pockets depending on the position of the Fermi level, have been reported in earlier works by analysing the Shubnikov-de Haas (SdH) oscillation for magnetic field along one of the crystallographic axes \cite{li,zwang,yli}. Due to the lower  symmetry (monoclinic) of crystal structure of TaSb$_{2}$, one expects a strong anisotropy in Fermi surface of this system.  However, the crystallographic direction dependence of magnetotransport properties and the anisotropic nature of the Fermi surfaces have not been probed.  This is important not only for understanding different electronic properties controlled by the Fermi surface but also helpful for the application point of view. In the present work, we have shown large anisotropy in magnetoresistance when the magnetic field is applied along different crystallographic directions in transverse experimental configuration (field is perpendicular to the current direction). Besides this, employing magnetization measurements along three mutually perpendicular directions on the same single crystal and by analysing the de Haas-van Alphen (dHvA) oscillation, we report the anisotropic nature of the Fermi pockets.
\section{Single crystal preparation, characterization and experimental details}
\begin{figure}[h!]
\includegraphics[width=0.45\textwidth]{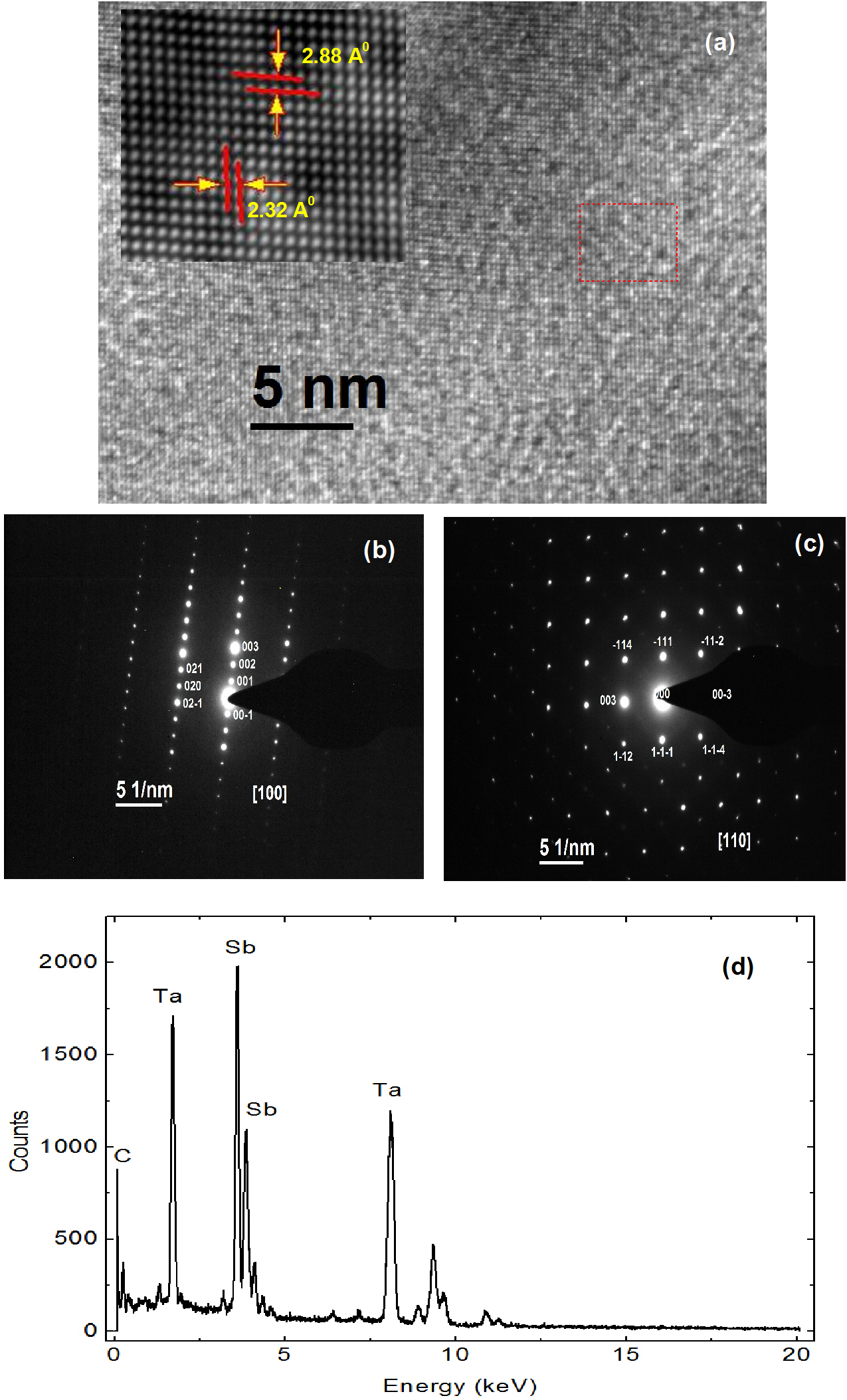}
\caption{(Color online) (a) High resolution TEM image, taken on a representative piece of TaSb$_{2}$ single crystals. Inset shows the Fourier-filtered image of the red dotted region. (b) and (c) are the selected area electron diffraction (SAD) patterns taken along [001] and [110] zone axis, respectively. (d) The energy-dispersive X-ray (EDX) spectroscopy data.}\label{rh}
\end{figure}
\begin{figure}[h!]
\includegraphics[width=0.45\textwidth]{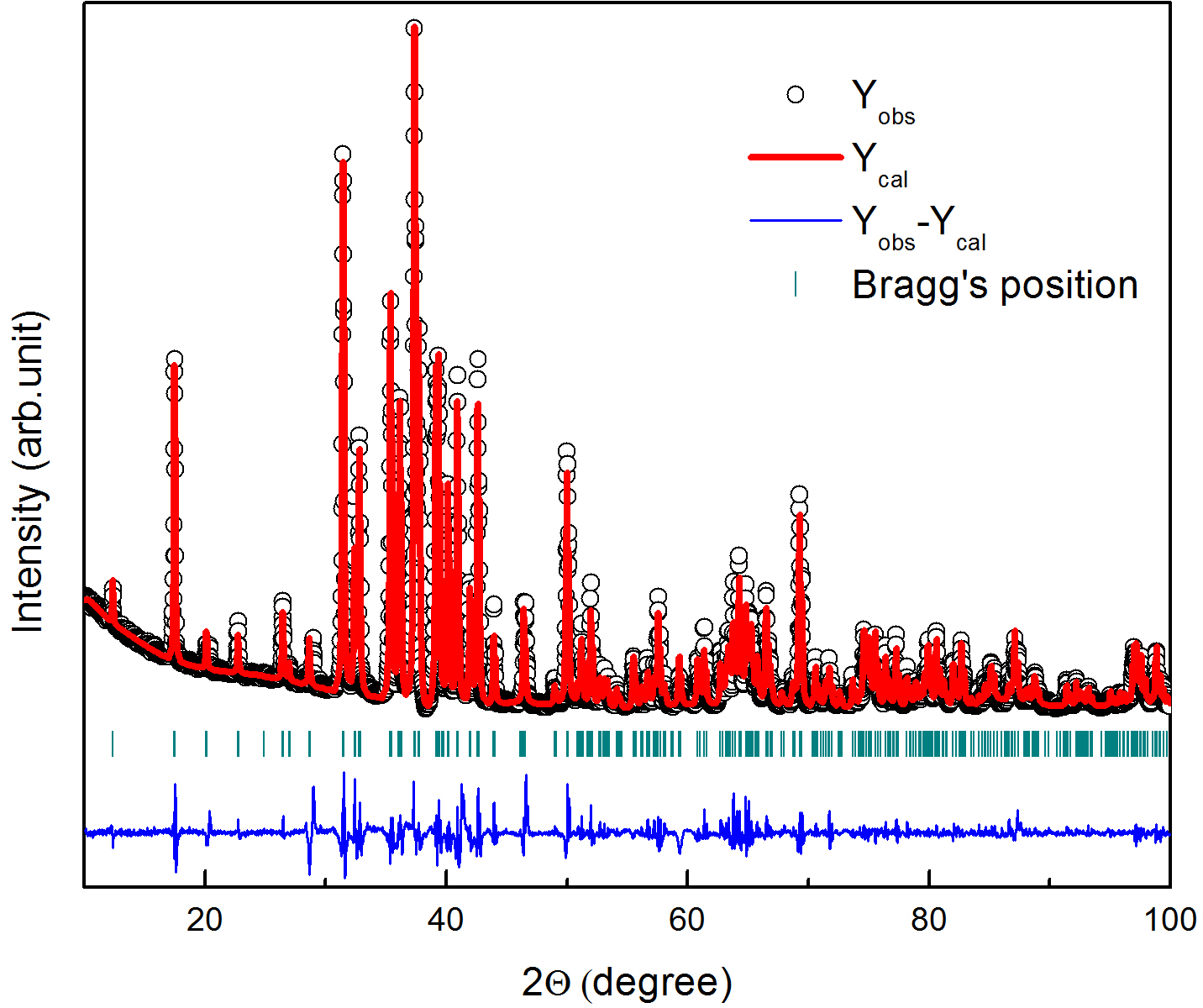}
\caption{(Color online) X-ray diffraction pattern of powdered single crystals of TaSb$_{2}$. Black open circles are experimental data (Y$_{obs}$), red line is the calculated pattern (Y$_{cal}$), blue line is the difference between experimental and calculated intensities (Y$_{obs}$-Y$_{cal}$), and green lines show the Bragg positions.}\label{rh}
\end{figure}
Single crystals of TaSb$_{2}$ were grown using iodine vapor transport technique in two steps. At first, polycrystalline sample is prepared by heating the stoichiometric mixture of high-purity Ta powder and Sb pieces at 650$^{\circ}$C for 8 h and at 750$^{\circ}$C for 48 h in a vacuum-sealed quartz tube. Finally, the quartz tube was placed in a gradient furnace and heated for 7 days. During heating, the end of the quartz tube containing the sample was maintained at 1000$^{\circ}$C,  while the other end was kept at  900$^{\circ}$C. The furnace was then cooled slowly to room temperature. Several small, shiny and niddle-like crystals formed at the cold end of the tube were mechanically extracted for transport and magnetic measurements. Phase purity and the structural analysis of the samples were done by using both the high resolution transmission electron microscopy (HRTEM) in FEI, Tecnai G$^{2}$ F30, S-Twin microscope operating at 300 kV equipped with energy dispersive x-ray spectroscopy (EDS, EDAX Inc.) unit  and the powder x-ray diffraction (XRD) technique with Cu-K$_{\alpha}$ radiation in a Rigaku x-ray diffractometer (TTRAX II). The HRTEM image of a representative piece of sample, which has been taken from a single crystal of TaSb$_2$  is shown in Fig. 1(a). Very clear periodic lattice structure implies that there is no secondary phase or atom clustering or disorder in the present sample. The Fourier-filtered image of the selected region in the inset, shows inter-planar spacings ($d$-spacing) of 2.88 {\AA} and 2.32 {\AA}. These measured $d$-spacings are close to the (111), and (003) inter-planar spacings of TaSb$_2$ (JCPDS \# 65-7656). Figure 1(b) and Figure 1(c) show the selected area electron diffraction (SAD) pattern recorded along [001] and [110] zone axis, respectively. The periodic pattern of the spots in SAD implies high-quality single crystalline nature of the grown samples. The diffraction pattern was indexed using the lattice parameters of monoclinic TaSb$_2$. The energy-dispersive x-ray (EDX) spectrum, as shown in Fig. 1(d), confirms the presence of the elements in desired stoichiometry. Please note that the carbon peaks in spectrum appear from the carbon coated grid on which the sample was mounted for TEM analysis. Figure 2 shows the high-resolution x-ray diffraction pattern of the powdered sample of TaSb$_{2}$ crystals at room temperature. Within the resolution of XRD, we did not see any peak due to the impurity phase.  Using the Rietveld profile refinement, we have calculated the lattice parameters $a$$=$10.221, $b$$=$3.645 and $c$$=$8.291 {\AA}, and $\beta$$=$120.40$^{\circ}$ with space-group symmetry $C_{12/m1}$. The transport measurements on TaSb$_{2}$ single crystals were done using the standard four-probe technique in a 9 T physical property measurement system (Quantum Design). The magnetization was measured in a 7 T MPMS3 (Quantum Design).
\section{Results}
\begin{figure}[h!]
\includegraphics[width=0.45\textwidth]{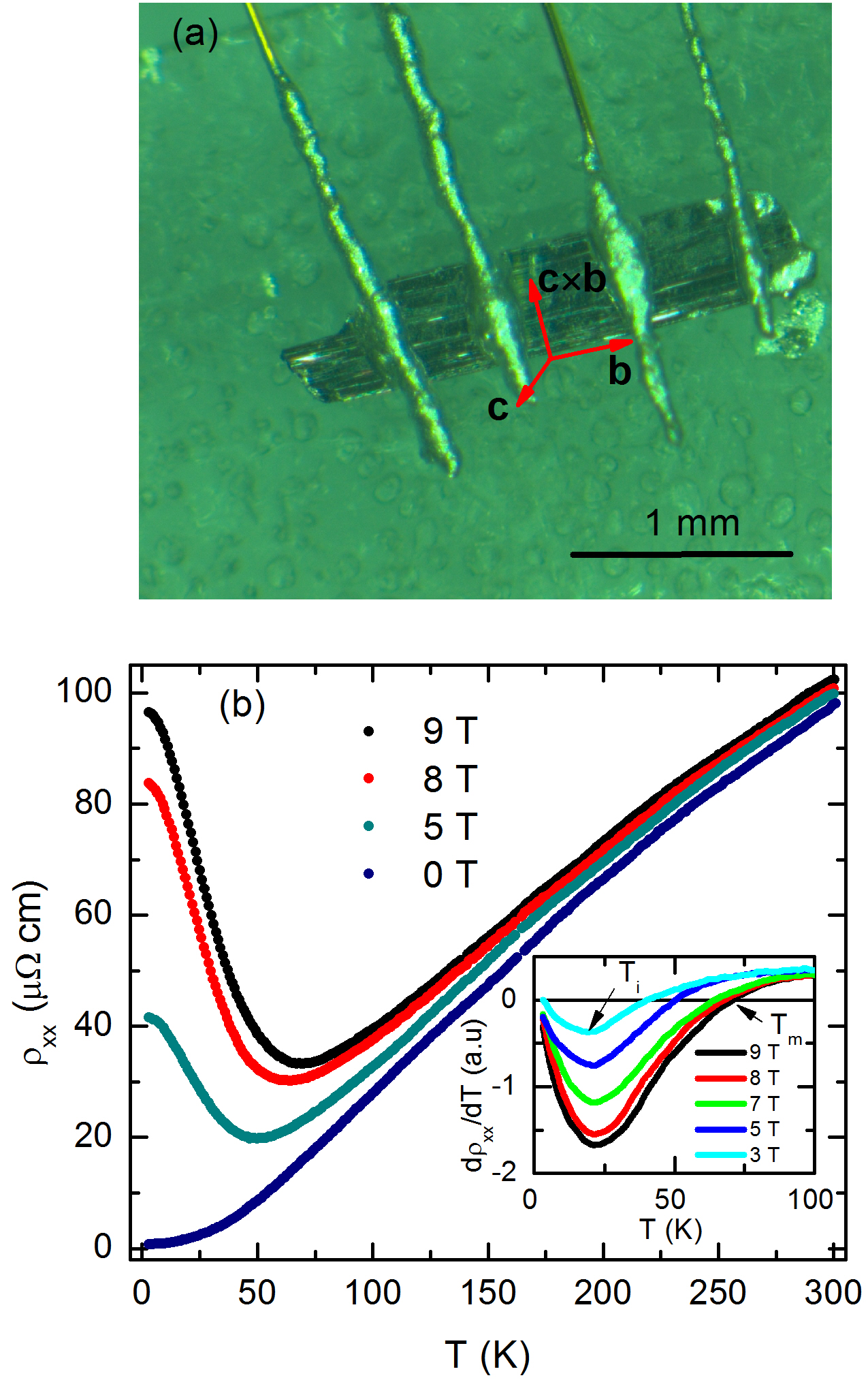}
\caption{(Color online) (a) Typical morphology and different crystallographic directions of a representative single crystal of TaSb$_{2}$, and (b) Temperature dependence of resistivity ($\rho_{xx}$) both in presence and absence of external magnetic field. Current ($I$) is applied along \textbf{b} axis and magnetic field ($H$) is along \textbf{c} axis. Inset shows the first order derivative of $\rho_{xx}$ with respect to $T$. Metal to semiconductor crossover temperature is named as $T_{m}$ and $T_{i}$ is the plateau temperature. }\label{rh}
\end{figure}
Figure 3(a) shows a representative single crystal of TaSb$_{2}$ with four electrical contacts. The typical length of the single crystals is $\sim$2 mm. This type of material, known as transition metal dipnictide MPn$_2$ [M = V, Nb, Ta, Cr, Mo, and W, Pn = P, As, and Sb], grows preferentially along \textbf{b} axis. As a consequence, the longer direction of the crystal is the \textbf{b} axis \cite{kwang,zyuan}. The largest flat plane has been found to be perpendicular to (001) direction \cite{zyuan}. Because of the monoclinic structure of the material, crystallographic \textbf{a} axis is not perpendicular to \textbf{c} axis. For convenience, we have defined three mutually perpendicular directions on the crystal as reference. Two of them are the crystallographic \textbf{c} and \textbf{b} axis, and the third one is perpendicular to \textbf{bc} plane, i.e., \textbf{(c$\times$b)} direction. The zero-field resistivity ($\rho_{xx}$) is metallic over the whole temperature range, as shown in Fig. 3 (b). $\rho_{xx}$ shows strong $T$ dependence. Small value of $\rho_{xx}$ at 2 K ($\sim$ 0.75 $\mu$$\Omega$ cm) and the large residual resistivity ratio, $\rho_{xx}$(300 K)/$\rho_{xx}$(2 K)$\sim$ 130, indicate good quality of the single crystals. With the application of magnetic field, the low-temperature resistivity drastically enhances. As a result, a metal to semiconductor-like crossover behaviour starts to appear with decreasing temperature. With the increase in field strength, the semiconducting-like behaviour becomes more and more prominent, and the metal to semiconductor like crossover temperature ($T_{m}$) shifts towards higher temperature side, as evident from the inset of Fig. 3 (b). At low temperature, $\rho_{xx}(T)$ shows a saturation like behaviour, similar to that observed in three-dimensional topological insulator at zero field. However, in the latter case, the saturation in $\rho_{xx}(T)$ appears due to competition between insulating bulk and metallic surface state. Unlike $T_{m}$, the temperature ($T_{i}$) at which $d\rho_{xx}/dT$ exhibits a minimum is almost independent of the strength of the magnetic field and remains fixed at $\sim$ 20 K. Slightly below $T_i$, the saturation-like behavior in $\rho_{xx}(T)$ starts to appear. The magnetic field induced metal-semiconductor crossover and the low-temperature resistivity plateau are the common phenomena in topological semimetals \cite{li,zwang,mnali,shekhar,yue,yzhao,tafti}. The actual microscopic origin of these phenomena are under debate. Different explanations such as magnetic field induced gap opening at the Dirac node \cite{shekhar,tafti} and Kohler's scaling of magnetoresistance \cite{ylwang} have been proposed.\\
\begin{figure}[h!]
\includegraphics[width=0.45\textwidth]{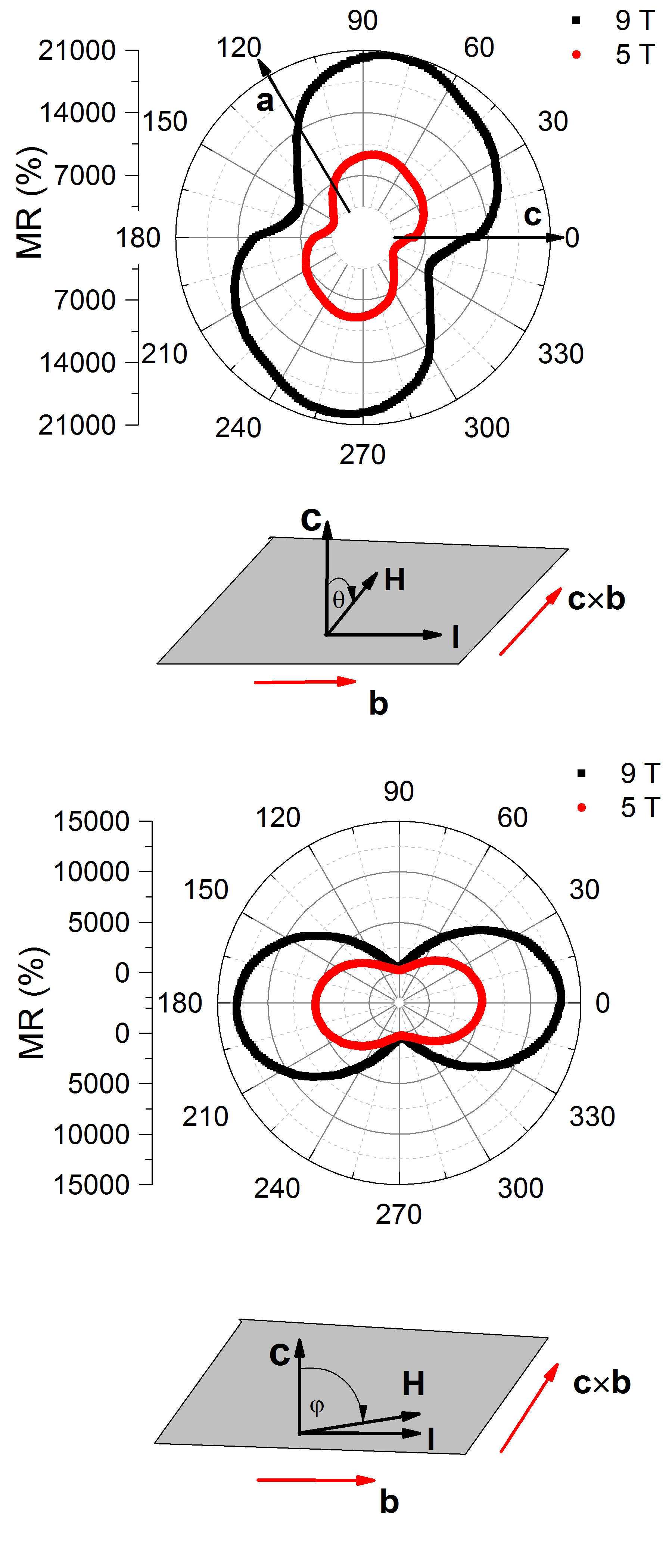}
\caption{(Color online) (a) Anisotropic magnetoresistance of a representative TaSb$_{2}$ single crystal at 2 K when the direction of magnetic field changes from \textbf{c}-axis to \textbf{(c$\times$b)} direction, making an angle $\theta$. (b) Magnetoresistance of the same piece of single crystal when the angle ($\phi$ - 90$^{\circ}$) between $I$ and $H$ has been changed at two representative field strength 5 and 9 T.}\label{rh}
\end{figure}

Several topological semimetals like NbSb$_{2}$, ZrSiS and TaAs$_{2}$ show anisotropic magnetoresistance with respect to the field direction, which arises due to the anisotropic nature of their Fermi surface \cite{kwang,ali1,zyuan,rsingha,jhu,yluo,rsing}. The anisotropic nature of magnetoresistance has huge impact in technological application and device fabrication. Figure 4(a) shows the transverse magnetoresistance ($H$ $\perp$ $I$) for the TaSb$_{2}$ single crystal at 2 K with the rotation of field  about \textbf{b} axis. When the field is along the \textbf{c} axis, the magnetoresistance (MR), which is defined as [$\rho_{xx}(H)$-$\rho_{xx}(0)$]/$\rho_{xx}(0)$, is $\sim$ 1.3$\times$10$^{4}$\% at 9 T and $\sim$ 4$\times$10$^{3}$\% at 5 T. As the direction of the field is changed from \textbf{c} axis towards \textbf{(c$\times$b)} direction, the value of MR is observed to increase and becomes maximum ($\sim$ 2$\times$10$^{4}$\% at 9 T) at around $\theta$ = 75$^{\circ}$. MR is minimum $\sim$ 9500\% at around 165$^{\circ}$. The polar plot in Fig. 4(a) shows a two-fold rotational symmetry, which is consistent with the monoclinic crystal structure of the present sample. The tilted pattern of MR($\theta$) with respect to the crystallographic axis may be due to complex geometry of the Fermi surfaces and their relative contribution to transport \cite{zyuan,coll}.\\

Figure 4(b) shows the typical behaviour of MR when the angle (90$^{\circ}$-$\phi$) between $I$ and $H$ has been varied continuously. As expected, due to the orbital origin of MR, the maximum and minimum in MR appear at $\phi$ = 0$^{\circ}$ and $\phi$ = 90$^{\circ}$, respectively. This also confirms that there is no intrinsic misalignment between between $I$ and $H$ in our crystal. Within the resolution of the angular variation of horizontal sample rotator, we have not observed any detectable negative MR under $H$$\parallel$$I$ configuration, i.e., in longitudinal set up. To further verify, we have also measured the field dependence of MR at small angle interval ($\sim$ 1$^{\circ}$) around $\phi$ = 90$^{\circ}$. We have repeated the same experiment on other single crystals but failed to detect any negative MR for $\phi$ close to 90$^{\circ}$. Li \emph{et al.} \cite{li} have reported negative longitudinal MR in TaSb$_{2}$. However, in a non-magnetic electronic system, there are mainly three possible origins of negative LMR known to exist.  In three-dimensional Dirac/Weyl systems, the most desirable one is due to the non-conservation of chiral charge, induced by externally applied magnetic field ($H$) parallel to electric field ($E$), known as Adler-Bell-Jackiw anomaly \cite{niel,quing}. It has been established that this negative LMR has a specific magnetic field and temperature dependence \cite{niel,quing,pariari}. The `current jetting' effect can be a possible microscopic origin of negative LMR, which has been found to appear from highly non-uniform current distribution inside the sample \cite{jing}. The strength and nature (i.e., field dependence) of this MR varies from sample to sample and with the location of the voltage leads. Besides the above two phenomena, the negative LMR has been found to observe in a few ultraclean layered materials such as PdCoO$_{2}$, PtCoO$_{2}$ and Sr$_{2}$RuO$_{4}$ \cite{kiku}. In these materials, negative LMR appears, when \textit{\textbf{E}} and \textit{\textbf{H}} are along certain crystallographic direction. For other directions, LMR is positive, unlike Dirac and Weyl semimetals. Moreover, in these systems, MR decreases linearly with increasing field from its zero-field value. So, to unambiguously establish the exact microscopic mechanism of negative LMR in a non-magnetic system, one has to look several factors such as the field dependence of $\rho$ and it's evolution with temperature, and needs to perform measurements for different contact configurations and along different crystallographic directions.\\

Figure 5(a) shows the field dependence of MR for $H$$\parallel$\textbf{c} and $I$$\parallel$\textbf{b} configuration. The large non-saturating MR up to 9 T suppresses with increasing temperature. Over the entire field range, it shows a typical $\sim$ $H$$^{1.5}$ dependence. Below 5 K, a high frequency Shubnikov-de Haas  effect has been observed in the high field region. Due to small amplitude and its rapid suppression with temperature, we have employed the de Haas-van Alphen  oscillation in magnetization measurements to probe the Fermi surface. Figure 5(b) shows the MR \emph{vs} $H$ for $H$$\parallel$\textbf{(c$\times$b)} and $I$$\parallel$\textbf{b} configuration. The value of MR is larger in this direction as evident from Fig. 4(a), and suppresses rapidly with increasing temperature. Before proceed to magnetization measurements, we have measured the Hall resistivity ($\rho$$_{xy}$) to estimate the value of carrier density ($n$) and mobility ($\mu$). As shown in Fig. 6, over the whole temperature and field range $\rho$$_{xy}$ is negative and approximately linear in $H$ at room temperature. With decreasing temperature, a weak nonlinearity appears in $\rho$$_{xy}(H)$ at high field due to presence of more than one Fermi pocket. From the low-field linear approximation of $\rho$$_{xy}(H)$, $n$ is calculated to be $\sim$ 3.1$\times$10$^{20}$ cm$^{-3}$ at 2 K. A quite large value of $\mu$ ($\sim$ 2.7$\times$10$^{4}$ cm$^{2}$/Vs at 2 K) has been obtained from the relation, $\mu$=[$\rho$$_{xx}$(0)$n$$e$]$^{-1}$.\\
\begin{figure}[h!]
\includegraphics[width=0.45\textwidth]{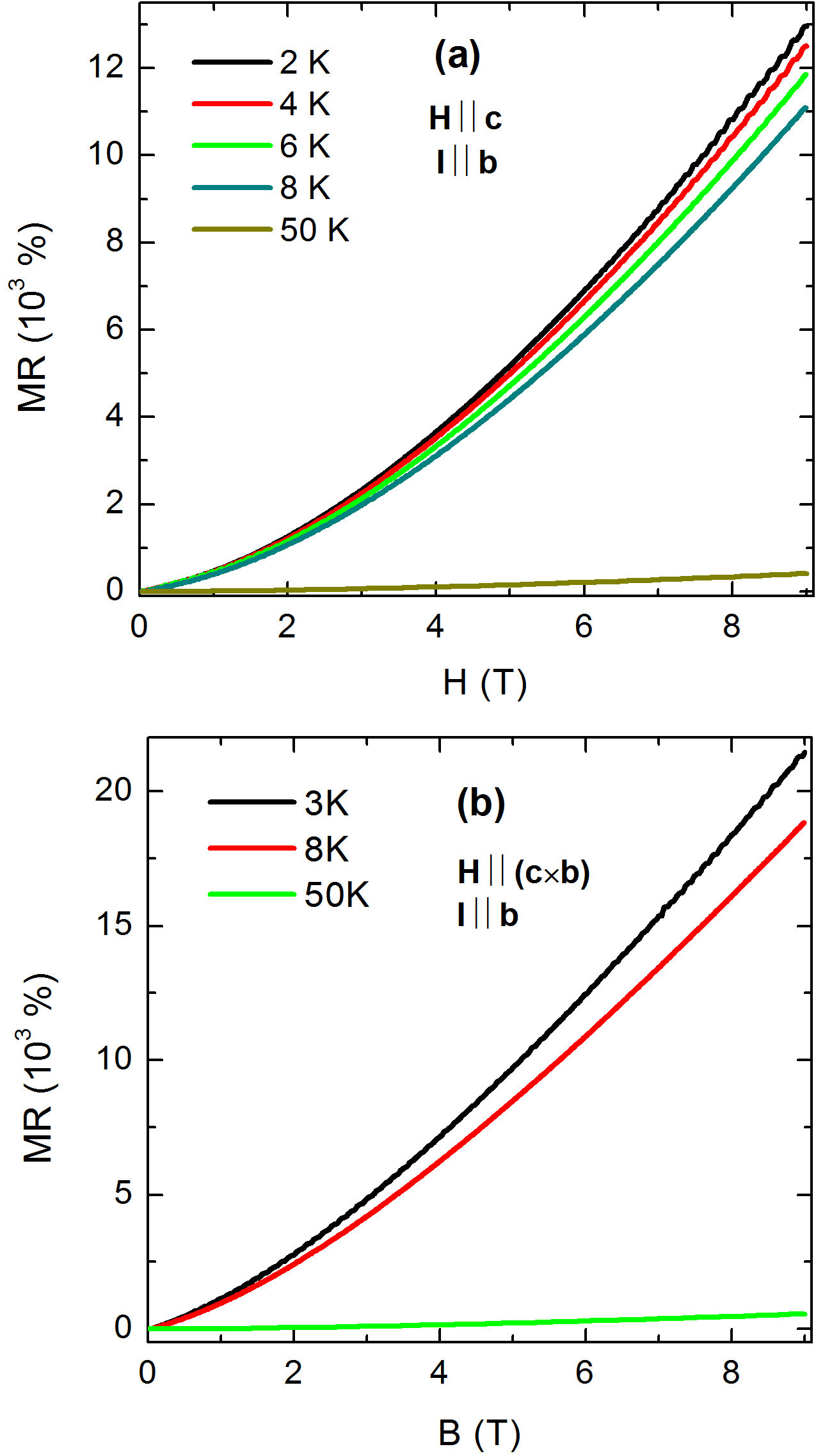}
\caption{(Color online) (a) Magnetoresistance as a function of magnetic field at some representative temperatures for (a) $H$$\parallel$\textbf{c} and $I$$\parallel$\textbf{b} configuration, and  (b) for $H$$\parallel$\textbf{(c$\times$b)} and $I$$\parallel$\textbf{b} configuration.}\label{rh}
\end{figure}

\begin{figure}[h!]
\includegraphics[width=0.45\textwidth]{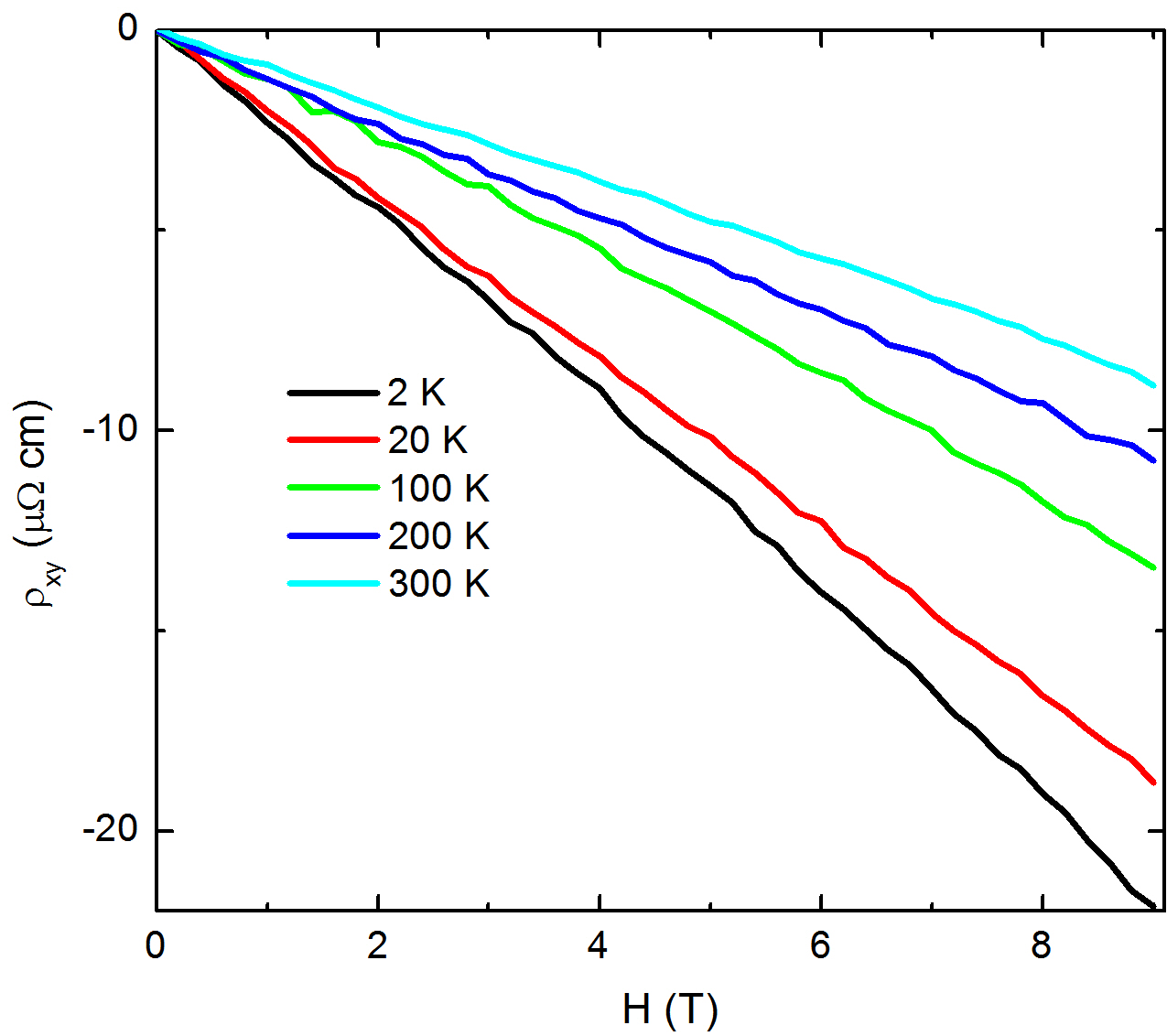}
\caption{(Color online) The field dependence of Hall resistivity ($\rho$$_{xy}$) and its evolution with temperature.}\label{rh}
\end{figure}

\begin{figure}[h!]
\includegraphics[width=0.45\textwidth]{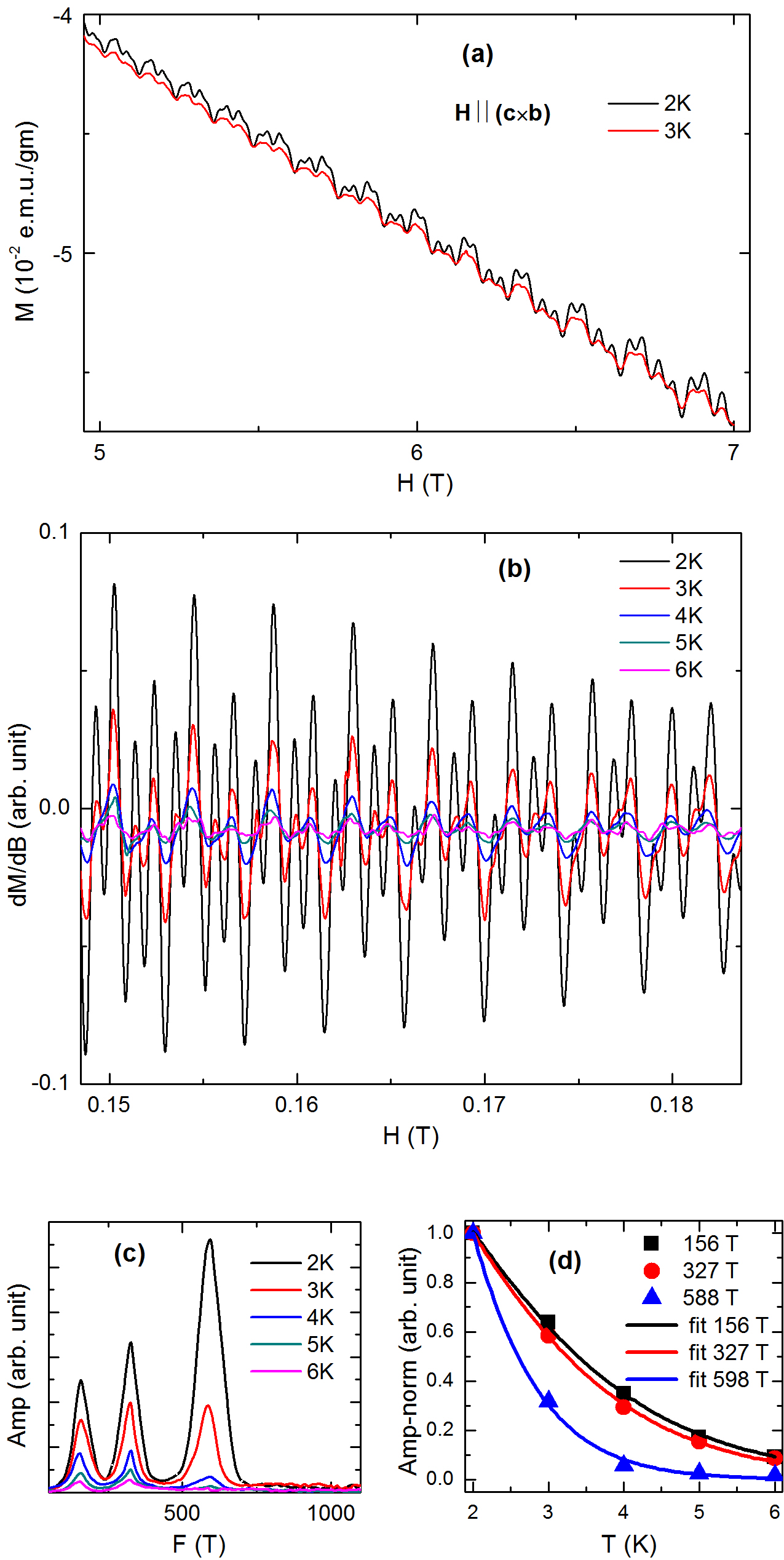}
\caption{(Color online) (a) Magnetic field (along \textbf{(c$\times$b)} direction) dependence of the diamagnetic moment of the TaSb$_{2}$ single crystal at two representative temperatures, 2 and 3 K. (b) Oscillating part of the susceptibility, obtained by taking the first order derivative of magnetization, i.e., $\Delta\chi$$=$ $d$($M$)/$d$$B$ versus 1/$B$. (c) The oscillation frequencies after the fast Fourier transformation. (d) Temperature dependence of the relative amplitude of the quantum oscillation. The solid line is a fit to the Lifshitz-Kosevich formula [Eq. (1)].}\label{rh}
\end{figure}
\textbf{TABLE I:} Parameters associated to the Fermi surface of TaSb$_{2}$, when the field is applied along the \textbf{(c$\times$b)} direction. $A_F$ is the Fermi surface cross-section. $m_{eff}$ is the effective mass of the charge carrier and $v_{F}$ is the Fermi velocity.
\begin{center}
\begin{tabular}{||c c c c||}
\hline
Frequency   & $A_F$ & $m_{eff}$ & $v_{F}$\\
T   & 10$^{-1}$${\AA}^{-2}$ & $m_{0}$ & 10$^{5}$m/s\\
\hline
156 & 14.6 & 0.32 & 2.5 \\

327 & 30.6 & 0.35 & 3.3 \\

598 & 56.3 & 0.59 & 2.6 \\
\hline
\end{tabular}
\end{center}
Figure 7(a) shows the magnetization  of TaSb$_{2}$ single crystal as a function of magnetic field ($H$) in $H$$\parallel$\textbf{(c$\times$b)} configuration. A distinct periodic pattern has been observed over the diamagnetic background, which can be interpreted as the superposition of more than one de Haas-van Alphen oscillation frequencies. Taking the first order derivative of the magnetization with respect to $H$, the oscillating component of the susceptibility has been obtained and shown in the Fig. 7(b). Although the oscillation is traceable down to 3 T, a smaller portion of $d$($M$)/$d$$B$ versus 1/$B$ plot has been shown to clearly visualize the pattern. From the figure, it is evident that the oscillation amplitude rapidly suppresses with increasing temperature and above 6 K, the amplitude is too small to detect within the experimental field range. The fast Fourier transform spectrum in Fig. 7(c) shows three distinct oscillation frequencies ($F$) at 156, 327 and 598 T. This implies that the present sample hosts three Fermi pockets similar to that observed in earlier SdH oscillation measurement \cite{zwang}. Three Fermi pockets are also expected from the theoretical calculation, for electron doped sample of TaSb$_{2}$ in electronic band structure \cite{zwang,yli,xu}. Employing the Onsager relation $F$$=$($\phi$$_0$/2$\pi$$^2$)$A_F$, the cross-sectional areas of the Fermi surface normal to the field direction have been calculated and listed in TABLE I. A significant difference in the cross section area of the Fermi pockets has been observed along this direction. The damping of the  oscillation amplitude with temperature can be described by the thermal damping term of the Lifshitz-Kosevich formula:
\begin{equation}
 \Delta R_{T}=a\frac{2\pi^2k_BT/\hbar\omega_c}{\sinh(2\pi^2k_BT/\hbar\omega_c)},
\end{equation}
where $a$ is a  temperature-independent constant  and $\omega_c$ is the cyclotron frequency. Figure 7(d) shows the fitting of the oscillation amplitude as a function of temperature with Eq. (1). Using the extracted value of $\omega_{c}$ from the fitting, the effective cyclotron mass of the charge carrier ($m_{eff}$) and  the Fermi velocity ($v_F$) are obtained from the relations $\omega_{c}$$=$$eB/m_{eff}$ and $v_F$$=$$\hbar$$k_F/m_{eff}$, respectively. The calculated parameters are shown in TABLE I. The effective mass of the charge carrier for all the three Fermi pockets are smaller than the rest mass of free electron and similar to that reported in earlier SdH oscillation study \cite{zwang}. Higher value of $m_{eff}$ for 588 T frequency implies that the band dispersion in this Fermi pocket is more non-linear in nature compare to the others.\\
\begin{figure}[h!]
\includegraphics[width=0.45\textwidth]{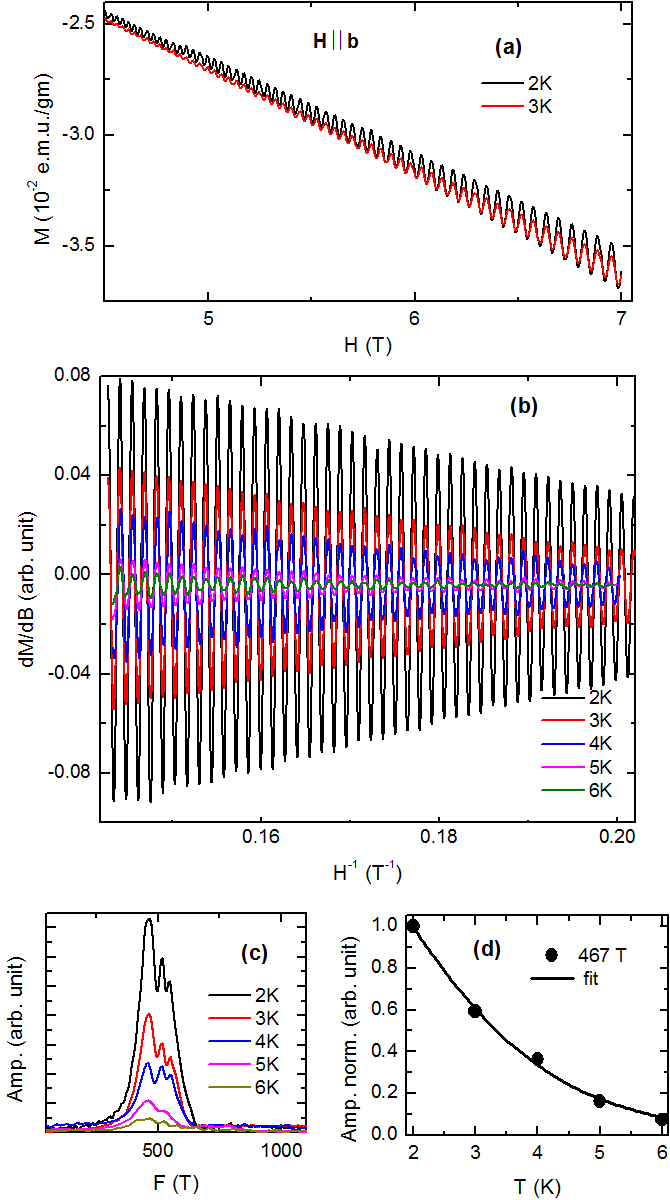}
\caption{(Color online) (a) Magnetic moment of the single crystal as a function of magnetic field at representative temperature of 2 and 3 K for $H$$\parallel$\textbf{b} configuration. (b) Oscillating part of the susceptibility $\Delta\chi$$=$ $d$($M$)/$d$$B$ versus 1/$B$. (c) The oscillation frequencies after the fast Fourier transformation. (d) Thermal damping of the relative oscillation amplitude. The solid line is a fit to the Lifshitz-Kosevich formula [Eq. (1)].}\label{rh}
\end{figure}

\textbf{TABLE II:} Parameters associated to the Fermi surface of TaSb$_{2}$, when the field is applied along the \textbf{b}-axis direction.
\begin{center}
\begin{tabular}{||c c c c||}
\hline
Frequency   & $A_F$ & $m_{eff}$ & $v_{F}$\\
T   & 10$^{-1}$${\AA}^{-2}$ & $m_{0}$ & 10$^{5}$m/s\\
\hline
467 & 44.4 & 0.33 & 4.2 \\

514 & 48.2 & - & - \\

545 & 51.4 & - & - \\
\hline
\end{tabular}
\end{center}
As the transverse magnetoresistance shows high anisotropy with respect to the field direction, we have performed dHvA oscillation study by applying field along other two directions of the crystal to probe the anisotropy in Fermi surface. Figure 8(a) shows the magnetization of the same piece of TaSb$_{2}$ single crystal when the field is applied along \textbf{b}-axis. Employing the fast Fourier transformation of the oscillations in Fig. 8(b), three closely spaced frequency peaks 467, 514  and 545 T have been obtained [Fig. 8(a)]. The calculated values of $A_F$ in TABLE II reveal nearly equal cross section area of Fermi pockets. If we compare these values with that observed in \textbf{(c$\times$b)} configuration, the overall cross section area is much larger along the present direction. As these frequency peaks are not well resolved, it will not be wise to calculate $m_{eff}$ and $v_F$ for each frequency. Only for the prominent peak at 467 T, we have calculated $m_{eff}$ and $v_F$ and shown in TABLE II. Although the value of $m_{eff}$ is close to that obtained in $H$$\parallel$\textbf{(c$\times$b)}, $v_F$ appears to be higher. Similar magnetic measurements and analysis have been done for the field along \textbf{c} crystallographic direction and shown in Fig. 9. This implies that the cross section areas of the Fermi pockets are comparable to each other, like $H$$\parallel$\textbf{b} configuration. The values of $m_{eff}$ and $v_F$ for the prominent peak at 372 T and $A_F$ for all the three frequencies are presented in TABLE III.\\

Employing magnetization measurements and by analyzing the dHvA oscillations, we have been able to obtain three distinct frequency peaks associated to three Fermi pockets  along three mutually perpendicular directions.  The three dimensional geometry of the Fermi surface can be constructed by observing the continuous evolution of the frequency peak associated to a particular Fermi pocket through extensive magnetization measurements at a small angle interval between crystallographic direction and magnetic field. At this moment, such type of set up is beyond our reach. However, the anisotropy in the Fermi surface and as a consequence, the anisotropy in the magnetoresistance of TaSb$_{2}$ single crystal is evident from the present study. Transverse magnetoresistance of orbital origin is scaled with the mobility of the charge carriers in the plane perpendicular to the external magnetic field \cite{coll,zengwei}. As mobility is a tensor quantity under application of magnetic field, the anisotropic nature of magnetoresistance can be explained by the anisotropy in the mobility tensor \cite{zengwei}. Again, the mobility ($\mu$) is determined by the ratio of the scattering time ($\tau$) to the effective mass of the charge carrier ($m_{eff}$), $\mu$ $\sim$ $\frac{\tau}{m_{eff}}$. In spite of the fact that we do not have complete knowledge on $m_{eff}$ for all the Fermi pockets, the larger magnetoresistance in $H\parallel$\textbf{(c$\times$b)} over $H\parallel$\textbf{c} configuration can be understood with the help of the scattering time ($\tau$). From the analysis of dHvA oscillation, we have seen  that the total cross-section area of the Fermi surfaces in the plane perpendicular to \textbf{(c$\times$b)} direction is smaller than the plane perpendicular to \textbf{c}-axis direction. This implies smaller phase space for the scattering of charge in the plane perpendicular to \textbf{(c$\times$b)} direction under application of magnetic field, and as a result, the value of $\tau$  is larger \cite{zengwei}. So, the mobility of the charge carriers in the plane perpendicular to \textbf{(c$\times$b)} appears to be higher. As the electrical conductivity is given by $\sigma(B)$ $\sim\frac{ne\mu}{1+\mu^2 B^2}$ \cite{jacob,coll,zengwei}, the value of magnetoresistance is expected to enhance with the rotation of field from \textbf{c} to \textbf{(c$\times$b)} direction. In the above discussion, we have not taken into account the effect of anisotropy in band mass ($m_{eff}$) on mobility tensor. However, it appears that the anisotropy in $m_{eff}$ is either weaker in competition with the anisotropy in $\tau$ or it has anisotropic behaviour similar to $\tau$. So, without considering the tensorial behaviour of $m_{eff}$, the anisotropic magneto resistance can be qualitatively explained.\\
\begin{figure}[h!]
\includegraphics[width=0.45\textwidth]{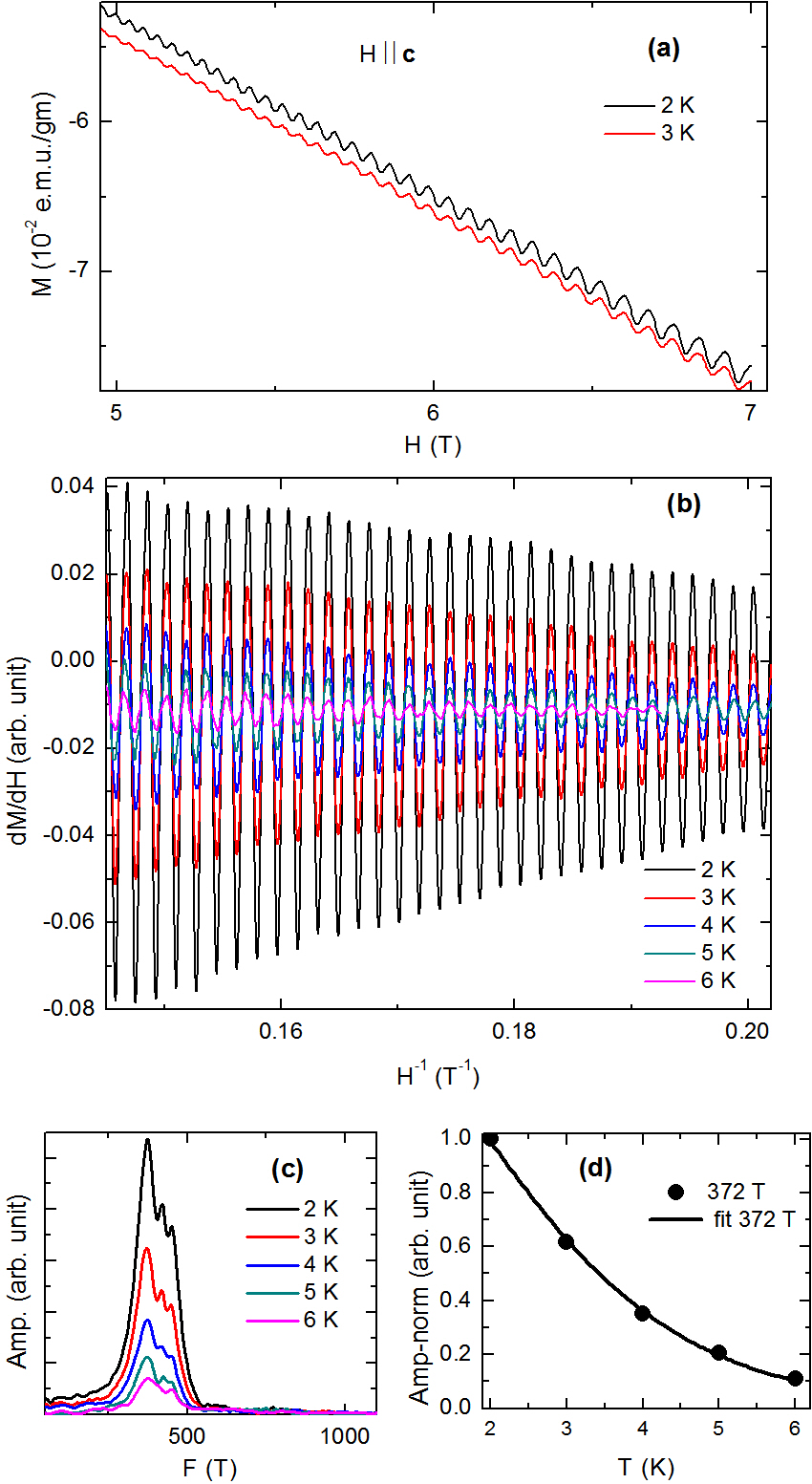}
\caption{(Color online) (a) Magnetic field dependence of the diamagnetic moment at representative temperature of 2 and 3 K for $H$$\parallel$\textbf{c} configuration. (b) Oscillating part of the susceptibility $\Delta\chi$$=$ $d$($M$)/$d$$B$ versus 1/$B$. (c) The oscillation frequencies after the fast Fourier transformation. (d) Thermal damping of the relative oscillation amplitude. The solid line is a fit to the Lifshitz-Kosevich formula [Eq. (1)].}\label{rh}
\end{figure}

\textbf{TABLE III:} Parameters associated to the Fermi surface of TaSb$_{2}$, when the field is applied along the \textbf{c}-axis direction.
\begin{center}
\begin{tabular}{||c c c c||}
\hline
Frequency   & $A_F$ & $m_{eff}$ & $v_{F}$\\
T   & 10$^{-1}$${\AA}^{-2}$ & $m_{0}$ & 10$^{5}$m/s\\
\hline
372 & 35.2 & 0.31 & 3.9 \\

417 & 39.3 & - & - \\

454 & 51.4 & - & - \\
\hline
\end{tabular}
\end{center}
\section{Conclusion}
In conclusion, we have observed a large anisotropy in the transverse magnetoresistance of TaSb$_{2}$ single crystal, by rotating the field along different crystallographic directions. Large nonsaturating magnetoresistance $\sim$ 20000\% at 9 T and 2 K reduces to $\sim$ 9500\% when the field is rotated from $\theta$ = 75$^{\circ}$ configuration towards $\theta$ = 165$^{\circ}$ configuration. Employing the magnetization measurements and analyzing the prominent de Haas-van Alphen oscillation within 7 T magnetic field, we have obtained three Fermi pockets as expected theoretically in electron doped sample of TaSb$_{2}$. Applying field along three mutually perpendicular directions of crystals, the cross sectional area of the Fermi pockets have been observed to vary. The angle dependence of transverse magnetoresistance can be understood by considering anisotropy in the quasi-particle scattering time.\\

\section{ACKNOWLEDGMENTS}
We thank Dr. Dilip Kumar Bhoi for his useful suggestion. We also thanks Arun Kumar Paul for his help during measurements.

\end{document}